# Analisis Keamanan Protokol Secure Socket Layer (SSL) Terhadap Proses Sniffing di Jaringan

Heru Pranata[1], Leon Andretti Abdillah[2], Usman Ependi[3]
[1,3] Program Studi Teknik Informatika, Fakultas Ilmu Komputer, Universitasa Bina Darma
[2] Program Studi Sistem Informasi, Fakultas Ilmu Komputer, Universitasa Bina Darma
Palembang, Indonesia
[1]jrs.herupranata@gmail.com, [2]leon.abdillah@yahoo.com

**Abstract.** Development of information technology, especially in the field of computer network allows the exchange of information faster and more complex and the data that is exchanged can vary. Security of data on communication in the network is a major thing. Secure socket layer (SSL) is the solution to the problem, but further research on the security of the SSL protocol transactions should be done to determine the extent of SSL can secure the data on the network. When the computer sends data across the network, the data is transmitted in packets. Sniffing is a technique of monitoring of every packet traversing the network. Security threat presented by sniffers is their ability to capture all incoming and outgoing packets through the network, which includes the passwords, usernames and other sensitive issues. Packet sniffer captures the data addressed to other devices, which will then be stored for later analysis later. Sniffing can also be used by system administrators to monitor the network and solve problems in the network.

## 1  Pendahuluan

Teknologi informasi (TI) tidak hanya memberikan sejumlah fasilitas yang canggih. Namun, TI juga ternyata memiliki sejumlah masalah yang berhubungan dengan keamanan. Tidak seorang pun ingin mengirim data pribadi mereka melalui internet kecuali mereka memiliki jaminan bahwa hanya penerima dimaksud yang akan menerimanya. Keamanan data pada komunikasi di jaringan merupakan hal utama yang bagitu penting. Komunikasi data melalui jaringan selalu mengandung kemungkinan kehilangan *confidentiality*, *message integrity* atau *endpoint authentication*. Terdapat tiga aspek utama ketika berbicara tentang keamanan data [1]: 1) *Privacy* atau *Confidentiality*, mencakup kerahasiaan informasi. Inti aspek *privacy* adalah bagaimana menjaga informasi agar tidak dilihat atau diakses oleh orang yang tidak berhak, 2) *Message Integrity*, atau integritas mencakup keutuhan informasi. Inti aspek *integrity* adalah bagaimana menjaga informasi agar tetap utuh atau dapat dikatakan apa yang diterima harus sama dengan apa yang dikirim, 3) *Authentication*, atau otentikasi berkaitan dengan keabsahan pemilik informasi. Harus ada cara untuk mengetahui informasi benar-benar asli, kemudian hanya yang berhak saja yang boleh memberikan informasi dapat dikatakan komunikasi harus dengan rekan yang tepat.

HTTPS adalah HTTP yang menggunakan *secure socket layer* (SSL). SSL adalah protokol enkripsi dipanggil melalui web server yang menggunakan HTTPS. SSL adalah jenis *sockets communications* berada diantara *transmission control protocol/internet protocol (TCP/IP)* dan *application layer*. SSL biasanya digunakan antara *server* dan *client* untuk mengamankan sambungan [2]. Akan tetapi studi lebih lanjut untuk menguji keamanan transaksi protokol SSL sendiri perlu dilakukan untuk melihat sejauh mana protokol SSL dapat mengamankan data di jaringan.

Ketika komputer mengirimkan data melalui jaringan, kemudian data tersebut dikirimkan dalam bentuk paket-paket. Paket *sniffing* adalah teknik pemantauan setiap paket yang melintasi jaringan. Paket *sniffing* adalah bagian dari perangkat lunak atau perangkat keras yang memonitor semua lalu lintas jaringan [3]. Komponen *sniffer* meliputi *hardware*, *drive program*, *buffer*, *packet analysis*. Ini tidak seperti jaringan *host* standar yang hanya menerima lalu lintas yang dikirim khusus untuk mereka. Ancaman keamanan yang disajikan oleh penyadapan adalah kemampuan mereka untuk menangkap semua lalu lintas masuk dan keluar, termasuk *password* dan *username* atau bahan sensitif lainnya [4].

## 2 Metode Penelitian

Metode yang digunakan pada penelitian ini adalah metode eksperimen. Didalam penelitian eksperimental, peneliti beraksi dari awal dalam hal pembentukan dan pemilihan kelompok, mencoba mengontrol, memutuskan apa yang terjadi pada setiap kelompok, mencoba mengontrol semua faktor lain yang relevan, mengontrol perubahan yang telah ia perkenalkan, dan pada akhir studi mengamati atau mengukur pengaruh atas kelompok-kelompok studi [5]. Penelitian ini terdiri atas beberapa tahap, yaitu: 1) Tahap Awal (Persiapan): a) Menentukan Objek/Subjek yang akan diteliti, b) Mengumpulkan teori-teori yang berhubungan dengan penelitian, c) Menentukan Variabel Penelitian, d) Membuat desain penelitian, 2) Tahap Pelaksanaan: a) Melaksanakan eksperimen/penelitian, b) Mengumpulkan data dari proses eksperimen, c) Menganalisis data, d) Menyusun laporan, 3) Tahap Akhir (Kesimpulan). Pada tahap ini terdapat gambaran mengenai hasil yang dari penelitian.

Pada penelitian ini digunakan tiga *tools sniffing*, yaitu: 1) *Wireshark* atau *Ethereal* [6] digunakan untuk *trouble shooting* jaringan, analisis, pengembangan software dan protokol serta untuk keperluan edukasi. *Wireshark* memungkinkan pengguna mengamati data dari jaringan yang sedang beroperasi atau dari data yang ada di *disk*, dan langsung melihat dan mensortir data yang tertangkap, 2) *SoftPerfect Network Analyzier*, alat profesional canggih untuk menganalisis, *debugging*, memelihara dan *monitoring* jaringan lokal dan koneksi *internet* [4, 7], dan 3) *Capsa 7 Free*. *Capsa* adalah *network analyzer* untuk LAN dan WLAN yang melakukan *real-time packet capture*, monitoring jaringan, protokol analisis, paket *decoding* mendalam, dan *expert* diagnosis otomatis [7].

Pada tahap pelaksanaan, penulis menggunakan aplikasi *web* berupa *blog*, *wordpress*. Menurut Abdillah [8] *weblog (blog)* sebagai media untuk membantu menyebarkan pengetahuan melalui *internet*. blog menjadi salah satu ikon paling populer dalam aplikasi internet saat ini. Blog memiliki banyak fitur yang mereka dapat sesuaikan.

## 3  Hasil dan Pembahasan

Hasil dengan dilakukannya analisis dan uji coba dalam penelitian ini menunjukkan pada HTTPS atau HTTP yang menggunakan SSL data dilindungi sebelum dikirim ke tujuan, perbedaanya sebelum menggunakan SSL cara kerjanya langsung mengirimkan data dalam bentuk *plaint-text* tanpa adanya perlindungan lebih. Pada SSL juga terdapat beberapa bagian protokol seperti, *handshake protocol*, *record protocol*, *alert*, dan *certificates*.

Protokol SSL memenuhi aspek keamanan berikut: 1) *Confidentiality*. Pada protokol SSL informasi sensitif yang dikirim antara *client* dan *server* bersifat rahasia, karena *plaint-text* dienkripsi mejadi *ciphertext* dengan bantuan *public key* dan *private key*, 2) *Message Integrity*. Peran *alert protocol* pada SSL adalah guna mencegah tindakan *intercept*, serta pengubahan pada pesan yang dikirim. Pada ujicoba yang dilakukan *message alert* terlihat ketika *client* melakukan pembatalan konfimasi *certificates* dan dimana *server* akan melakukan pembatalan komunikasi dengan *client*, dan *client* tidak dapat melakukan akses terhadap aplikasi *web* pada *server*, dan 3) *Authentication*. Peran *digital certificates* pada protokol SSL adalah untuk memastikan jalur komunikasi SSL yang dibangun hanya yang berhak saja yang boleh memberikan informasi atau dapat dikatakan komunikasi dilakukan dengan rekan yang tepat.

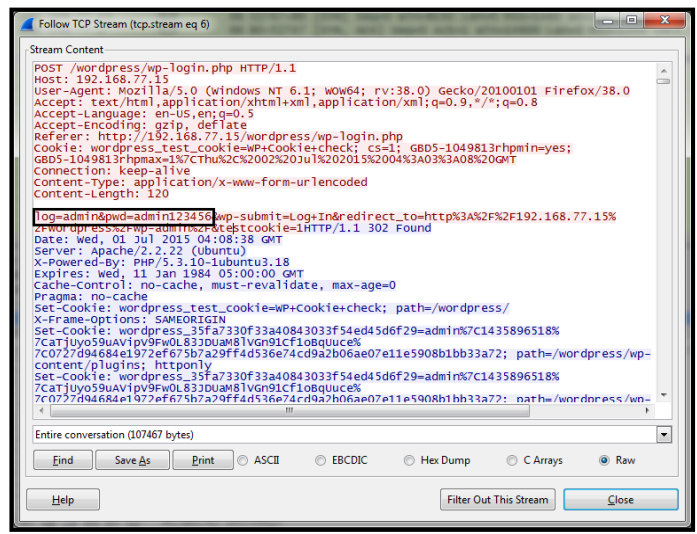

**Gambar 1.** Hasil *sniffing* sebelum menggunakan SSL

### 3.1 Test Awal (*Pre-Test*)

Peneliti melakukan tes awal (*pre-test*) pengujian dengan melakukan *sniffing* terhadap proses komunikasi yang terjadi antara *client* dan *server*. Dari hasil *sniffing* tersebut dapat dilihat protokol yang digunakan adalah HTTP karena belum menggunakan SSL (gambar 1).

Dalam perpindahan tersebut terdapat *delay* waktu. Jika menggunakan *ping time*, terdapat beberapa kali *Request Time Out* (RTO) sedangkan jika menggunakan *download* koneksi mengalami *drop* beberapa detik sebelum kembali berjalan normal, tergantung dari *server download*.

### 3.2 Test Akhir (*Post-Test*)

Setelah tahapan-tahapan konfigurasi SSL selesai dilakukan peneliti melakukan post-test guna melihat perbedaan yang terjadi sebelum dan sesudah diterapkanya SSL pada aplikasi web pada penelitian ini (gambar 2).

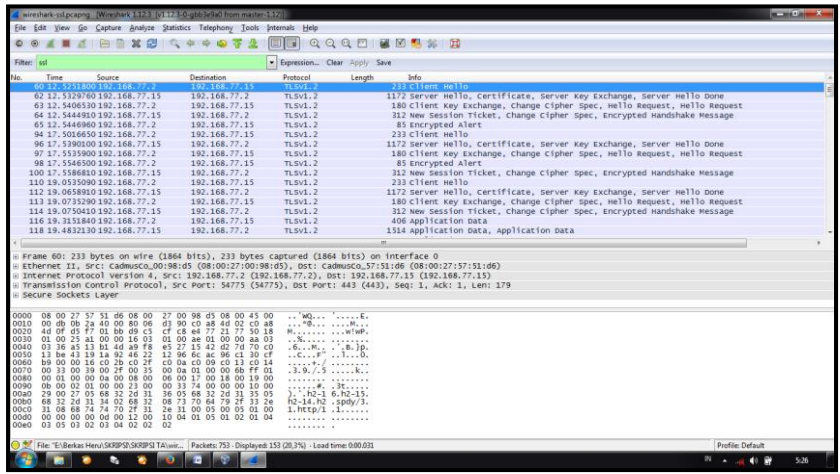

**Gambar 2.** Hasil *sniffing* setelah menggunalan SSL.

### 3.3 Analisis

Pada penelitian ini digunakan tiga *tools sniffing*, yaitu: 1) *Wireshark*, 2) *SoftPerfect Network Analyzier*, dan 3) *Capsa 7 Free*. Peneliti melakukan ujicoba dengan melakukan *sniffing* secara bersamaan menggunakan ketiga *sniffing tools* tersebut dan didapatkan perbandingan data hasil *packet capture* seperti pada tabel 1.

Selanjutnya peneliti melakukan analisis lebih lanjut dengan menggunakan data hasil *packet capture* tersebut dan melakukan analisis terhadap proses komunikasi SSL antara *client* dan *server*. Serta mencoba melakukan dekripsi terhadap data hasil *sniffing* tersebut.

Peneliti mencoba memanfaatkan log komunikasi SSL yang ada pada *browser* yang digunakan *client* dengan cara menangkap *pre-master secrets/master secrets* melalui *browser* dengan menggunakan *Environment Variables* dan memanfaatkan *log* tersebut untuk melakukan dekripsi menggunakan *wireshark*.

**Tabel 1.** Perbandingan Ukuran *Packet Capture*

| No. | Sniffing tools | Ukuran Packet Capture sebelum menggunakan SSL | Ukuran Packet Capture setelah menggunakan SSL |
|---|---|---|---|
| 1 | *Wireshark* | 416 kb | 460 kb |
| 2 | *SoftPerfect Network Analyzer* | 397 kb | 437 kb |
| 3 | *Capsa 7 Free* | 463 kb | 516 kb |

Gambar 3 memperlihatkan langkah-langkah mengatur *Environment Variables* (Windows 7) yang dimanfaatkan untuk menangkap *pre-master secrets/master secrets* selama proses komunikasi SSL antara *client* dan *server* dibangun.

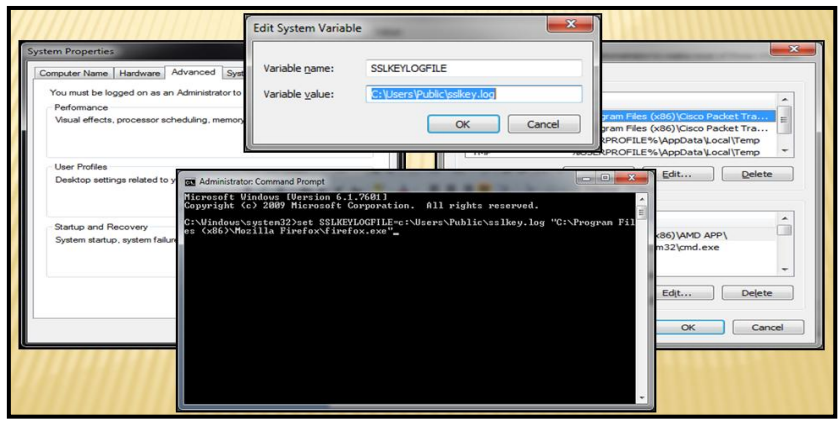

**Gambar 3.** *Setting Environment Variables*

Hasil dari *setting environment variables* di atas, peneliti mendapatkan *log* selama proses komunikasi SSL antara *client* dan *server* dibangun (gambar 4). Kemudian hasil *log* tersebut digunakan untuk melakukan dekripsi dengan menggunakan *Wireshark*.

Setelah peneliti melakukan analisis terhadap semua hasil *sniffing* pada saat *client* dan *server* melakukan pengiriman data pada jalur komunikasi SSL tersebut. Peneliti menemukan beberapa proses yang mampu didekripsi dengan menggunakan *key log file* pada *browser client*, namun tidak ditemukan informasi berupa *username* dan *password*.

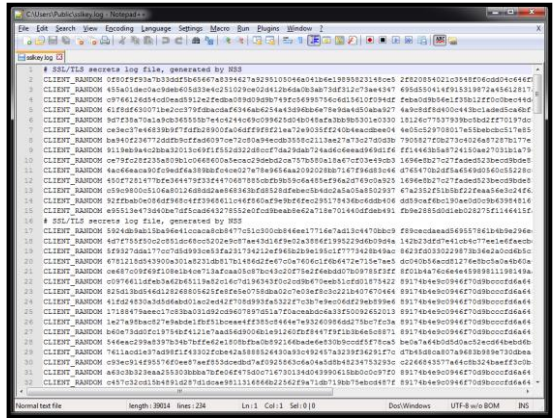

**Gambar 4.** *Key Log File*

## 4  Kesimpulan

Berdasarkan hasil ujicoba dengan melakukan analisis menggunakan penyadapan (*sniffing*) terhadap komunikasi data antara *client* dan *server*, maka dapat disimpulkan:
1. Protokol SSL merupakan protokol yang aman dari tindakan *sniffing*.
2. Penggunaan protokol SSL pada jaringan juga sangat penting guna mengamankan data di jaringan ini.

## Daftar Pustaka